\documentclass[preprint,superscriptaddress,preprintnumbers,
groupedaddress,
nofootinbib,
 amsmath,amssymb,
 aps]{revtex4-1}
\usepackage{multirow}
\usepackage{float}
\usepackage{latexsym}
\usepackage{graphicx}
\usepackage{amssymb}
\usepackage{amsmath}
\usepackage{amsfonts}
\usepackage[hidelinks]{hyperref}
\usepackage{color}
\usepackage{cool}
\usepackage{dcolumn}
\usepackage{slashed}
\usepackage{caption}
\usepackage{subcaption}

\begin{document}
\preprint{TUM-HEP-1263/20}

\title{Fat brane and seesaw mechanism in extra dimensions}

\author{Bj\"orn Garbrecht}
\author{Ricardo G. Landim}\email{ricardo.landim@tum.de}

\affiliation{Technische Universit\"at M\"unchen, Physik-Department,\\ James-Franck-Stra\text{$\beta$}e, 85748 Garching, Germany}


\date{\today}

\begin{abstract}
In this paper we present a higher-dimensional seesaw mechanism. We consider a single, flat extra dimension, where a fat brane is localized and contains the standard model (SM) fields, similar to Universal Extra Dimension  models.  There is only one Dirac fermion in the bulk, and in four dimensions it results  in two towers of Kaluza-Klein (KK) Majorana sterile neutrinos, whose mass mixing  with the SM neutrinos is suppressed due to a brane-localized kinetic term. The interaction between the sterile neutrinos and the SM is through the usual coupling with the Higgs boson, where the coupling depends upon the compactification radius $R^{-1}=10^{-2}-1$ GeV and the width of the fat brane $L^{-1}=2$ TeV, where the latter value is chosen to avoid LHC constraints.  Due to this suppression mechanism the mass of the lightest sterile neutrinos can be of order $\mathcal{O}(1-10)$~\text{TeV} while naturally explaining the small SM neutrino mass, which in turn is   easily obtained for a large range of parameter choices.   Furthermore, neutrino oscillations are not substantially influenced by the tower of sterile KK particles. Finally, leptogenesis is investigated in this setup, and it is viable for some values within the parameter space.
\end{abstract}

\maketitle

\section{Introduction}
Extra dimensions (ED) have been used to explain a plethora of phenomena in particle physics and cosmology, including the  hierarchy
 \cite{Antoniadis:1990ew,Dienes:1998vh,Antoniadis:1998ig,ArkaniHamed:1998rs,Randall:1999ee,Arkani-Hamed:2016rle,Arun:2017zap,Arun:2018yhg} and flavor problems \cite{Agashe:2004cp,Huber:2003tu,Fitzpatrick:2007sa}, proton stability \cite{Appelquist:2001mj},  the origin of electroweak symmetry breaking \cite{ArkaniHamed:2000hv,Hashimoto:2000uk,Csaki:2002ur,Scrucca:2003ut}, the breaking of grand unified gauge groups \cite{Hebecker:2001jb,Hall:2001xr,Asaka:2002my,Asaka:2003iy},  the number of fermion
generations \cite{Dobrescu:2001ae,Fabbrichesi:2001fx,Borghini:2001sa,Fabbrichesi:2002am,Frere:2001ug,Watari:2002tf}, the seesaw mechanism \cite{Dienes:1998sb,Huber:2003sf},   and leptogenesis \cite{Pilaftsis:1999jk}. The standard model (SM) itself can be enlarged if its content is  promoted to fields that propagate into a compact ED, in the so-called Universal Extra Dimension (UED) models. In this scenario, the  zero-mode of each Kaluza-Klein (KK) state is seen in four dimensions as the correspondent SM particle. UED models were built in 5-D \cite{Appelquist:2000nn} and 6-D \cite{Dobrescu:2004zi,Burdman:2005sr,Ponton:2005kx,Burdman:2006gy}, whose compactification radius $L$ is constrained using supersymmetry searches at the LHC \cite{Aad:2015mia},  since both 
models can have  similar phenomenology.  The current bound for the 5-D UED model is $L^{-1}>1.4-1.5$ TeV 
\cite{Deutschmann:2017bth,Beuria:2017jez,Tanabashi:2018oca} (for $\Lambda L \sim 5-35$, where $\Lambda$ is
the cutoff scale), while for the 6-D  UED model the bound  is $L^{-1}>900$ GeV \cite{Burdman:2016njl}.

When a brane is present, kinetic terms  can be induced in it as the result of loop corrections associated with the interaction between the fields in the bulk and localized matter fields in the brane. The resulting induced brane-localized kinetic term  (BLKT) describes a massless field, being effectively 4-D for distances shorter than the compactification radius. Such a mechanism was studied in 5-D for spin-2 field \cite{Dvali:2000rx}, gauge theories \cite{Dvali:2000hr} and supersymmetric models \cite{Hebecker:2001ke,Cheng:2002ab}, giving also  similar results in 6-D   \cite{Dvali:2000xg,Dvali:2001ae,Dvali:2002pe}. Additionally, the localization of matter or gauge fields in branes has been explored in other contexts, for thin \cite{ArkaniHamed:1998rs,Dvali:1998pa, Alencar:2014moa,Alencar:2017dqb,Alencar:2015awa,Alencar:2015rtc,Alencar:2015oka,Alencar:2018cbk,Freitas:2018iil,Fichet:2019owx} and thick branes  \cite{DeRujula:2000he,Georgi:2000wb}, while BLKT has been investigated in  different scenarios \cite{Carena:2002me,Carena:2002dz,delAguila:2003gv,delAguila:2003bh,Davoudiasl:2002ua,Davoudiasl:2003zt,Datta:2012xy,Datta:2013nua,Dey:2013cqa,Dey:2015pba,Dey:2016cve,Dasgupta:2018nzt,Chiang:2018oyd, Rizzo:2018ntg,Rizzo:2018joy}.

 In a recent paper \cite{Landim:2019epv}, an ED was employed along with a finite width `fat' brane to explain the expected smallness of the coupling between SM and a dark mediator, where this mediator is either a vector or a scalar field. In this setup, a fat brane was used and the interaction between the SM and the mediators was found to be suppressed, when compared with the coupling between these same mediators and a dark matter candidate, confined in a separate thin brane. The suppression mechanism is much more accentuated when a BLKT is present. A similar result was obtained for a vector field in the bulk, in a model with two ED \cite{Landim:2019ufg,Landim:2019gdj}. 
 
The small value of the SM neutrino mass can be explained if one uses the seesaw mechanism, where the diagonalization of the neutrino mass matrix leads to a massive mostly sterile neutrino and a very light mostly active neutrino. This mechanism is known to be possible using large extra dimensions \cite{Dienes:1998sb} or warped geometry \cite{Huber:2003sf}, so that a natural extension of previous works \cite{Landim:2019epv} would be to investigate if a fermion in the bulk, playing the role of a sterile neutrino, has its interaction with SM suppressed in such a way that the seesaw mechanism can be realized. This is the purpose of the present work. We find that the two towers of Majorana sterile neutrinos can indeed provide the explanation of the small SM neutrino mass through a higher-dimensional seesaw mechanism.  The  lightest sterile neutrino masses can be of order $\mathcal{O}(1- 10)$ TeV because the mass mixing between the sterile neutrinos and the SM neutrino is naturally very suppressed for a wide range of parameter choices.
Neutrino oscillations are not influenced by the tower of sterile neutrinos since the survival probability is practically equal to one. Finally, we investigate leptogenesis in this setup, showing that it can explain the observed baryon asymmetry of the universe for some values of the parameter space.

Since with the present scenario, for a type-I seesaw framework at the TeV-scale, we aim to explain the observed smallness of the masses of the light neutrinos through the suppression of their Yukawa couplings employing an ED with a fat brane and a BLKT, it is distinct from other important realizations of the seesaw mechanism at low scales. In particular, the so-called inverse seesaw mechanism~\cite{Wyler:1982dd,Mohapatra:1986bd,Mohapatra:1986aw,Mohapatra:2016twe,Branco:1988ex,GonzalezGarcia:1988rw,Kersten:2007vk,Abada:2007ux,Gavela:2009cd} relies on an approximately conserved lepton number that is extended to the right-handed neutrinos. This leads to a suppression of the lepton-number violating Majorana masses of the light, mostly active neutrinos, while leaving relatively large mixing angles with the heavier, mostly sterile neutrinos. A distinct feature is therefore the enhanced mixing
of active and sterile neutrinos that may even lead to the observation of the heavy neutrino states with current and planned experiments.
A particularly interesting variant of such models is based on the left-right symmetric model (LRM) (see Ref.~\cite{Mohapatra:2016twe} for a review), where intriguing phenomenology arises from the additional Higgs mechanism at the TeV scale. In contrast, in the present model, the new states are feebly coupled sterile neutrinos that currently appear impossible to be observed in combination with the rich phenomenology implied by the presence of the ED.
 
 This paper is organized as follows. In Sect. \ref{sec:bulk} we consider a fermion in the bulk and derive the equations of motion and wave functions. The interaction with the SM is presented in Sect. \ref{sec:seesaw}  along with the seesaw mechanism and the survival probability of the SM neutrino. In Sect. \ref{sec:leptogenesis} we investigate leptogenesis and Sect. \ref{sec:conclu} is reserved for conclusions.

\section{Neutrino in the bulk}\label{sec:bulk}

We consider a single, flat ED represented by an interval $0\leq y\leq \pi R$, with a fat brane localized between $\pi r$ and $\pi R$, where the SM is confined, and we assume $L\equiv R-r\ll R$.
There is only one generation of a Dirac fermion in the bulk  $\Psi(x^\mu,y)$, with Dirac and Majorana mass terms.  We consider the induced kinetic term in the fat brane, so that the corresponding action is \cite{Huber:2003sf,delAguila:2003bh}
\begin{equation}
    S=\int d^4 x\, dy\Big[i \Bar{\Psi}\Gamma^A\partial_A  \Psi -m_D \Bar{\Psi} \Psi - m_M \Bar{\Psi} \Psi^c +\mathcal{L}_{BLKT}\Big]\,,
\end{equation}
where $A=0-3, 5$ is the 5-D index, $m_D$ is the Dirac mass, $m_M$ is the Majorana mass, $\Gamma^4=i\gamma^5$ and $\Psi^c= C^5 \Bar{\Psi}^T$ is the charge conjugated
spinor, with $C^5=\gamma^0 \gamma^2 \gamma^5$. The BLKT in the action is given by \cite{delAguila:2003bh}\footnote{For simplicity, we  restrict our attention  for the case where  the BLKT parameter $\delta_A$ is the same for the two components of the Dirac spinor $\psi_1$, $\psi_2$, \textit{i.e.} $\delta_A^1=\delta_A^2\equiv \delta_A$, although different contributions might be possible, as presented in \cite{delAguila:2003bh}.  }
\begin{equation}
    \mathcal{L}_{BLKT}=i \Bar{\Psi}\slashed{\partial}\Psi\cdot \delta_A \theta(y) R\,,
    \end{equation}
where the step-function is
\begin{equation}
    \theta(y)=0 \quad \text{for $y<\pi r$}, \quad \theta=\alpha \quad \text{for $\pi r< y \leq \pi R$}\,,
    \end{equation}
    with $\delta_A>0$ and  $\alpha $ being a positive constant with dimensions of energy. We define region I as $0\leq y<\pi r$ and region II as $\pi r< y\leq \pi R$. 
Writing the bulk sterile neutrino as $\Psi=\Psi_1+\Psi_2$,\footnote{There are no chiral fermions in 5-D and  chirality in 4-D is recovered through the $Z_2$ orbifold symmetry  $y\rightarrow -y$, where one spinor is taken to be even under this symmetry, while the second one is taken to be odd. As we shall see, the wave functions do not satisfy orbifold boundary conditions (BC) in the region $\pi r< y< \pi R$, therefore we do not obtain a chiral  spinor.  In addition, the Dirac mass term in 4-D would be canceled if one had used  orbifold BC, which again, is not the case here. }    we can expand it as a tower of KK states
\begin{equation}\label{decomposition}
    \Psi_{1,2}(x^\mu,y)=\sum_{n=0}^\infty f_{1,2}^{(n)}(y)\psi_{1,2}^{(n)}(x^\mu )\,.
\end{equation}

Using this decomposition the 4-D action is found after integrating out the ED, where the wave functions $f_{1,2}(y)$ satisfy the following orthogonality relations
\begin{equation}\label{normalization}
    \int_0^{\pi R} dy [1+\delta_AR \theta(y)]( f_{1}^{(m)}f_{1}^{(n)}+f_{2}^{(m)}f_{2}^{(n)})=\delta_{m,n}\,,
\end{equation}
\begin{equation}\label{massNorm}
    \int_0^{\pi R} dy f_{2}^{(m)}\partial_y  f_{1}^{(n)}=- \int_0^{\pi R} dy f_{1}^{(m)}\partial_y f_{2}^{(n)}\Big]=m_n\delta_{m,n}\,.
\end{equation}
In order for Eq. (\ref{massNorm}) to be true, the 
integration by parts gives the following
 coupled BC $f_1^{(m)}f_2^{(n)}(\pi R)-f_1^{(m)}f_2^{(n)}(0)=0$, which should be satisfied for all $m$ and $n$. 
With the decomposition (\ref{decomposition}) and using the Majorana condition for the 4-D fields $\Bar{\psi}_{1}^{(n)}=\psi_2^{(n)}$, we get the equation of motion for the two components of the wave function \cite{Huber:2003sf}
\begin{equation}\label{1stEDO}
    (\pm \partial_y-m_D)f_{1,2}^{(n)}+[(1+\delta_A R \theta(y))m_n-m_M]f_{2,1}^{(n)}=0\,.
\end{equation}
The first-order  equations (\ref{1stEDO}) can be transformed into a second-order  equation for $f_2^{(n)}$, for example.\footnote{The choice of which wave function would have a second-order equation is arbitrary. If we had chosen  $f_1^{(n)}$ instead of $f_2^{(n)}$, the final solution would have a overall minus sign. The transcendental equation (to be shown next) would still be the same. } This procedure gives
\begin{equation}\label{2ndEDO}
    \partial^2_y f_2^{(n)} +\big[(m_n-m_M+m_n\delta_A \theta(y) R)^2-m_D^2 \big] f_2^{(n)}=0\,,
\end{equation}
where $m_n=\sqrt{x_n^2/R^2+m_D^2}+m_M$ and the roots $x_n$ will be determined by the appropriate transcendental equation. Having determined $f_2^{(n)}$, the solution is then replaced in the respective Eq. (\ref{1stEDO}), to solve for $f_1^{(n)}$.

The solutions of Eqs. (\ref{1stEDO}) and (\ref{2ndEDO}) are found for the two different regions in the ED space, that is, inside the fat brane and outside it. The solution for the wave function in  region I have the form $f_2^{(n)}(y)= A_n \cos(x_n y/R )+B_n\sin(x_n y/R )$, but imposing that it vanishes at $y=0$ in order to satisfy the coupled BC, we have the following solutions for this region
\begin{align}
  f_{1,\text{I}}^{(n)}(y)&=\frac{\Lambda_{n}}{m_n-m_M}\Big[m_D\sin\left(\frac{x_n y}{R}\right)+\frac{x_n}{R} \cos \left(\frac{x_n y}{R}\right)\Big]\,, \\ 
  f_{2,\text{I}}^{(n)}(y)&=\Lambda_{n} \sin \left(\frac{x_n y}{R}\right)\,,
\end{align}
where $\Lambda_{n}$ is the normalization constant found using Eq. (\ref{normalization}). The wave functions in the region II have solutions of the form $f_{1,2,\text{II}}^{(n)}=A_n \cos(\bar m_n y)+B_n\sin(\bar {m}_n y)$, where the constants $A_n$ and $B_n$ for one of the components are determined using the condition of continuity of the function and  continuity of its derivative, at $y=\pi r$. The second wave function must satisfy Eq. (\ref{1stEDO}), thus the resulting solutions are \cite{Landim:2019epv}
\begin{align}
    f_{1,\text{II}}^{(n)}(y)&=f_{1,\text{I}}^{(n)}(\pi r)\cos[\Bar{m}_n(y-\pi r)]+\frac{f_{1,\text{I}}^{'(n)}(\pi r)}{\Bar{m}_n}\sin[\Bar{m}_n(y-\pi r)]\,,\label{f1II}\\
    f_{2,\text{II}}^{(n)}(y)&=\frac{m_n-m_M }{\sqrt{m_D^2+\Bar{m}_n^2}}\Lambda_n \sin \left (\frac{x_n \pi r}{R}\right) \cos[\Bar{m}_n(y-\pi r)]\nonumber\\&+\frac{\Bar{m}_n  f_{1,\text{I}}^{(n)}(\pi r)+f_{1,\text{I}}^{'(n)}(\pi r)m_D/\Bar{m}_n }{\sqrt{m_D^2+\Bar{m}_n^2}}\sin[\Bar{m}_n(y-\pi r)]\,,\label{f2II}
\end{align}
where $\Bar{m}_n^2\equiv (m_n-m_M+m_n\delta_A \alpha R)^2-m_D^2$ and the prime is a derivative with respect to $y$. 

Using Eq. (\ref{normalization}), the normalization constant is
\begin{align}
   2\Lambda_n^{-2}&= 2\pi r+(1+\delta_A\alpha R)\bigg[\frac{A_{1}^{(n)}B_{1}^{(n)} +A_2^{(n)}B_2^{(n)}}{\Bar{m}_n}+(A^{(n)2}_{1}+B^{(n)2}_{1}+A_2^{(n)2}+B_2^{(n)2})\pi L\nonumber\\
    &-\frac{A_{1}^{(n)}B_{1}^{(n)}+A_2^{(n)}B_2^{(n)}}{\Bar{m}_n}\cos(2\Bar{m}_n \pi L)+\frac{A^{(n)2}_{1}-B^{(n)2}_{1}+A_2^{(n)2}-B_2^{(n)2}}{2\Bar{m}_n} \sin(2 \Bar{m}_n \pi L)\bigg]\nonumber\\&-\frac{m_D[\cos(2\pi rx_n/R)-1]}{ x_n^2/R^2+m_D^2}+\frac{(x_n^2/R-m_D^2)\sin(2\pi r x_n/R))}{2(x_n^2/R^2+m_D^2)x_n/R}\nonumber\\
    &-\frac{\sin(2\pi rx_n/R)}{2 x_n/R}\,,
\end{align}
where  $A_{1(2)}^{(n)}$ and $B_{1(2)}^{(n)}$ are the terms (obviously without the normalization constant) that multiply the cosine and the sine in Eq. (\ref{f1II}) or Eq. (\ref{f2II}), respectively.

Imposing  $f_{1,\text{II}}^{(n)}=0$ to satisfy the remaining part of the coupled BC, we get the  transcendental equation that determines the roots $x_n$
\begin{equation}\label{transceq}
  \tan(\Bar{m}_n\pi L)=-\Bar{m}_n \frac{f_{1,I}^{(n)}(\pi r)}{f_{1,I}^{'(n)}(\pi r)}\,.
\end{equation}

The 4-D Lagrangian contains Dirac and Majorana mass terms, but we can form the linear combination $N_1^{(n)}=(\psi_1^{(n)}+\psi^{(n)}_2)/\sqrt{2}$ and $N_2^{(n)}=i(\psi_1^{(n)}-\psi^{(n)}_2)/\sqrt{2}$, such that they diagonalize the mass matrix. The resulting tower of Majorana eigenstates have the corresponding  physical masses given by $m_{1(2)}^{(n)}=\sqrt{x_n^2/R^2+m_D^2}\pm m_M>0$, where a hierarchy between the mass parameters and a positive  bulk Majorana mass is assumed to assure that  the physical masses are always positive. 

\section{Seesaw mechanism}\label{sec:seesaw}

\begin{figure}
    \centering
    \includegraphics[scale=0.45]{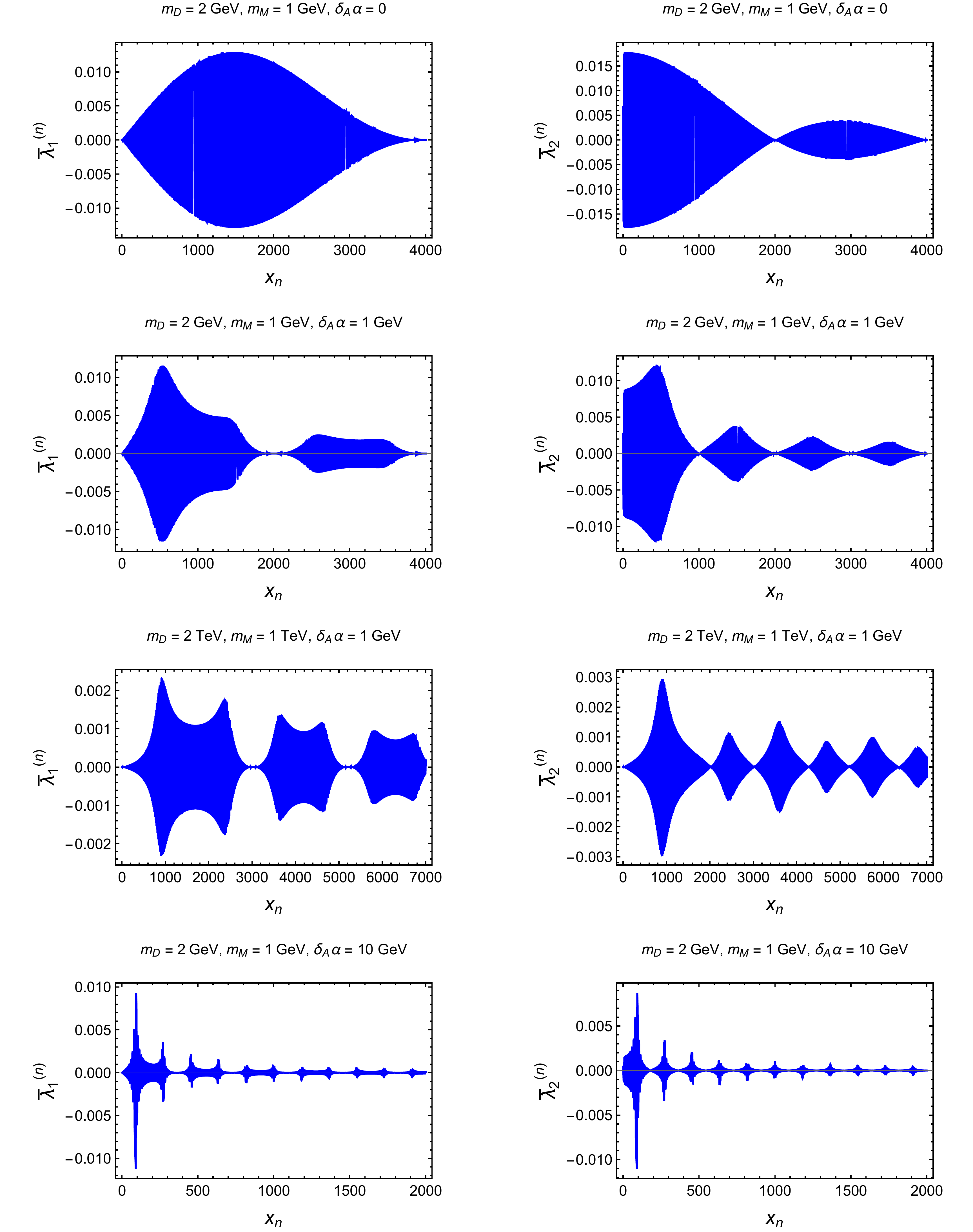}
    \caption{Oscillatory behavior of the couplings $\Bar{\lambda}_1^{(n)}$ (left) and $\Bar{\lambda}_2^{(n)}$ (right), for different values of $m_D$, $m_M$ and $\delta_A\alpha$, for $\lambda_{5,1(2)}=1$ TeV$^{-1/2}$, $R^{-1}=1$ GeV and $L^{-1}=2$ TeV. }
    \label{fig:plotBM0}
\end{figure}

\begin{figure}
    \centering
    \includegraphics[scale=0.45]{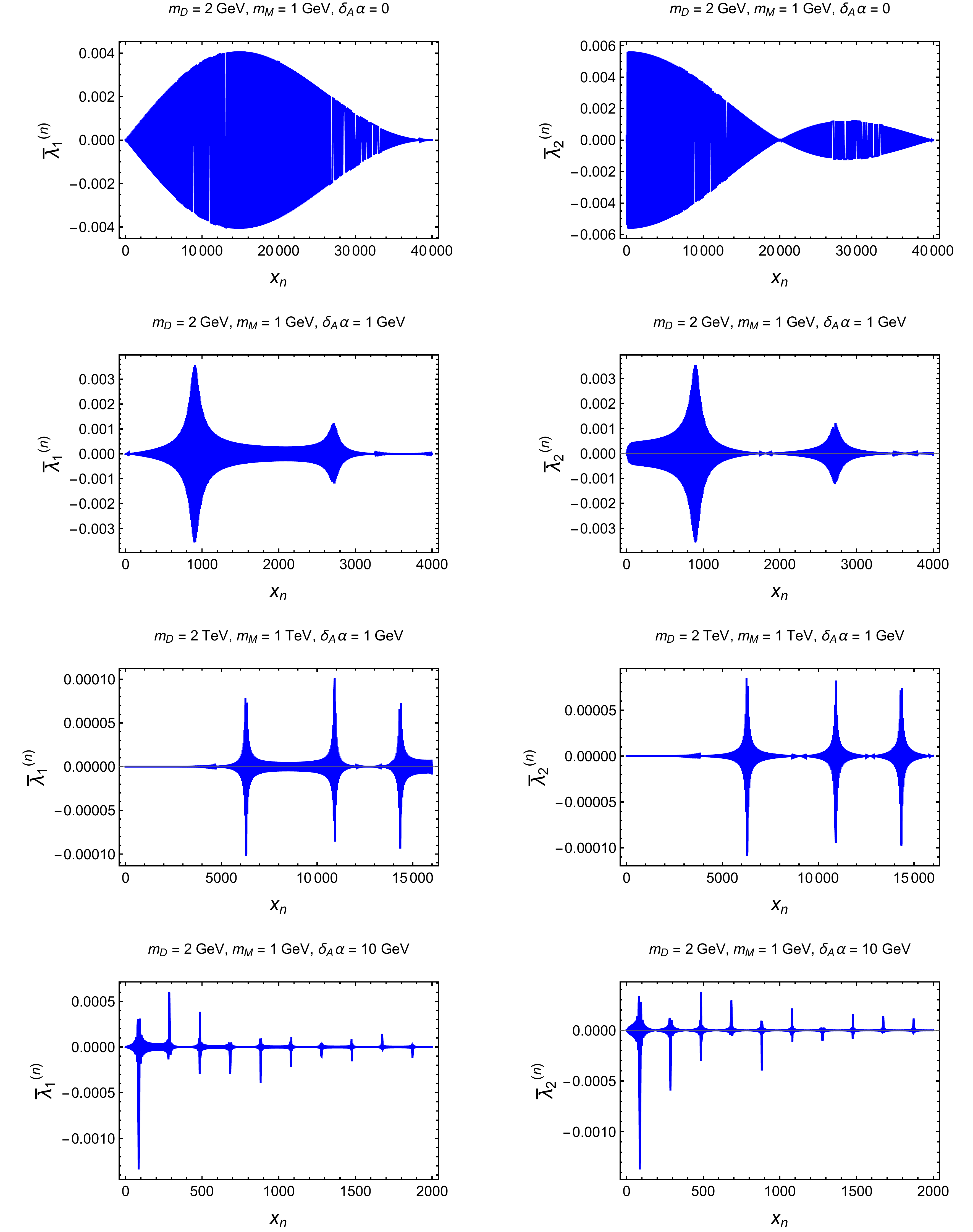}
    \caption{Oscillatory behavior of the couplings $\Bar{\lambda}_1^{(n)}$ (left) and $\Bar{\lambda}_2^{(n)}$ (right), for different values of $m_D$, $m_M$ and $\delta_A\alpha$, for $\lambda_{5,1(2)}=1$ TeV$^{-1/2}$, $R^{-1}=100$ MeV and $L^{-1}=2$ TeV. }
    \label{fig:plotBM1}
\end{figure}

\begin{figure}
    \centering
    \includegraphics[scale=0.45]{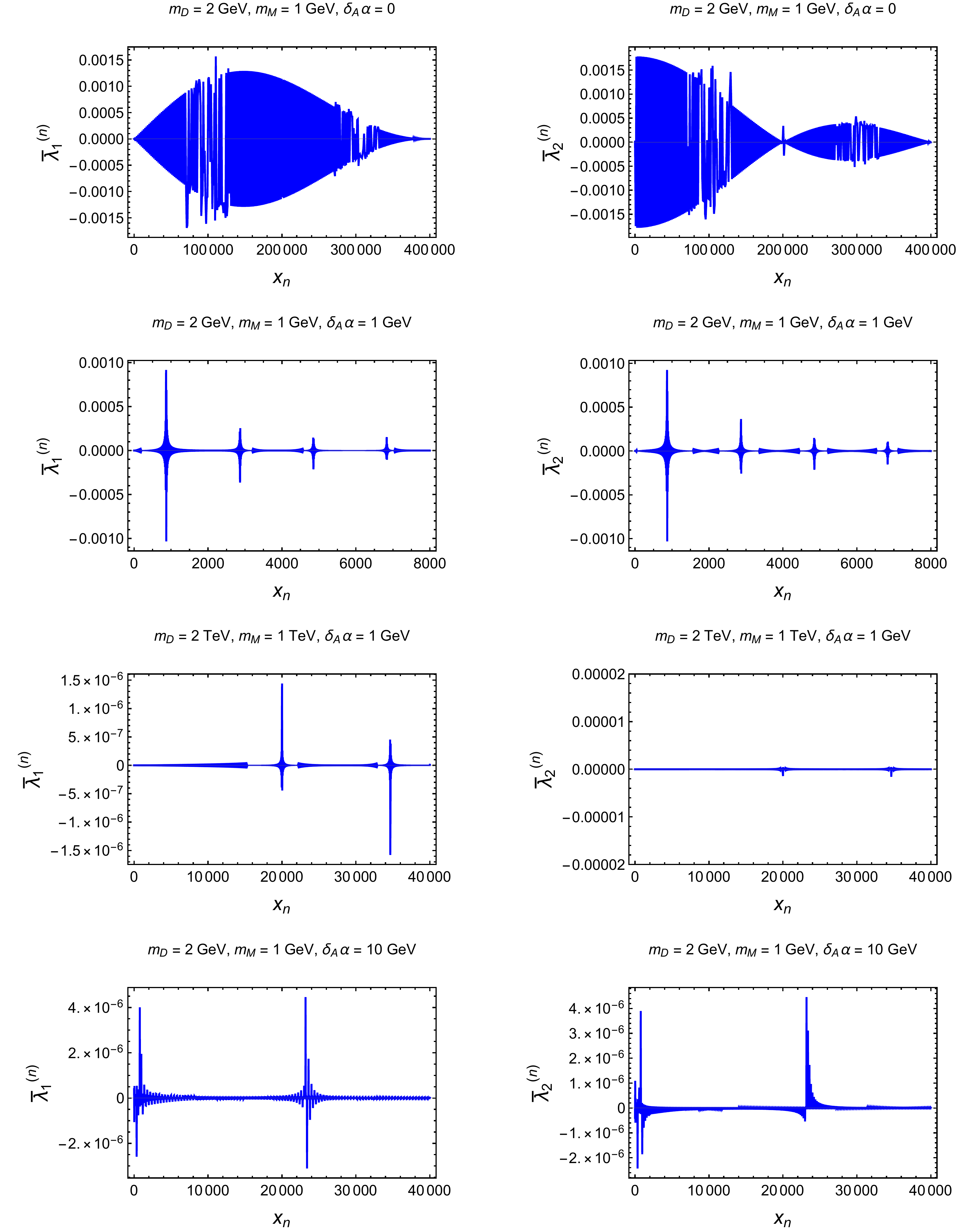}
    \caption{Oscillatory behavior of the couplings $\Bar{\lambda}_1^{(n)}$ (left) and $\Bar{\lambda}_2^{(n)}$ (right), for different values of $m_D$, $m_M$ and $\delta_A\alpha$, for $\lambda_{5,1(2)}=1$  TeV$^{-1/2}$, $R^{-1}=10$ MeV and $L^{-1}=2$ TeV. }
    \label{fig:plotBM2}
\end{figure}

There is an interaction between the bulk fermion, the Higgs $H$ and the $SU(2)_L$ doublet fermion $L_f$  given by $\lambda_5 \Psi L_fH+ \text{h.c.}=\lambda_{5,1} \Psi_{1} L_fH+\lambda_{5,2} \Psi_{2} L_fH+\text{h.c.}$, where $\lambda_5$ is the Yukawa matrix, $\lambda_{5,1(2)}$ are the correspondent  5-D Yukawa couplings and h.c. stands for Hermitian conjugation. We will omit flavor indices. Since we are interested in the interaction with conventional SM particles, we will assume only the zeroth-KK mode for the SM fields.  After expanding the sterile neutrino in a KK tower of states, the 4-D couplings $\Bar{\lambda}_{1(2)}^{(n)}$ are defined as
\begin{align}\label{eq:couplings}
\Bar{\lambda}_{1(2)}^{(n)}&\equiv \frac{\lambda_{4,1(2)}}{\Lambda_0} \int_{\pi r}^{\pi R} \,dy \frac{f_{1(2)\text{II}}^{(n)}(y)}{\pi L}\nonumber\\
&=\frac{\lambda_{4,1(2)}\Lambda_n}{\Bar{m}_n\pi L\Lambda_0} \biggr \{B_{1(2)}^{(n)}\Big[1-\cos (\Bar{m}_n \pi L)\Big]+A_{1(2)}^{(n)}\sin(\Bar{m}_n \pi L)\biggr\}\,,
\end{align}
where $\lambda_{4,1(2)}\equiv \lambda_{5,1(2)}\Lambda_0$  is defined to be a 4-D dimensionless Yukawa coupling,  $(\pi L)^{-1/2}$ is the usual  normalization of the UED SM fields and recall that $L$ is the width of the fat brane. We plot the couplings $\Bar{\lambda}_1^{(n)}$ and $\Bar{\lambda}_2^{(n)}$ as  functions of the roots $x_n$ in Figs. \ref{fig:plotBM0}--\ref{fig:plotBM2}, for different values of the parameters. We took $\lambda_{5,1(2)}=1$ TeV$^{-1/2}$  without loss of generality. From the figures we see that increasing  either the value of the bulk masses or $\delta_A\alpha$ decreases the couplings, while the case without BLKT in the fat brane ($\delta_A\alpha=0$) presents the least suppressed behavior.  The presence of BLKT, therefore, makes the coupling smaller and more suppressed, shrinking the oscillatory pattern as the combination $\delta_A\alpha$ is increased.  This is the same behavior found in  \cite{Landim:2019epv} (where the coupling was proportional to $L/R$, for the lightest KK states), where here we can also see that larger compactification radius $R$ decreases the couplings as well. 

The interaction in the mass eigenstate basis is $\lambda^{(n)}_{1(2)}N^{(n)}_{1(2)}L_fH+$ h.c., where $\lambda_{1}^{(n)}=(\Bar{\lambda}_1^{(n)}+\Bar{\lambda}_2^{(n)})/\sqrt{2}$ and $\lambda_{2}^{(n)}=-i(\Bar{\lambda}_1^{(n)}-\Bar{\lambda}_2^{(n)})/\sqrt{2}$.  We are ignoring  possible additional phases for the couplings, because it turns out that both couplings can be turned into real numbers by a phase adjustment, so that in what follows only the absolute value of them is important. 

After the spontaneous symmetry breaking via  the Higgs mechanism the off-diagonal mass term  $\frac{1}{2}\hat{m}_{1(2)}^{(n)}N_{1(2)}^{(n)}\nu_L+$ h.c. appears in the Lagrangian, where $\hat{m}_{1(2)}^{(n)}\equiv \sqrt{2}\lambda_{1(2)}^{(n)}v$ and $v=246$ GeV is the Higgs vacuum expectation value. The mass term for the  neutrinos can be written as $\frac{1}{2}\mathcal{N}^T\mathcal{M N}+$ h.c., where
\begin{equation}\label{eq:gauge eigenstates}
\mathcal{N}^T\equiv(\nu_L,N_1^{(0)},N_2^{(0)},N_1^{(1)},N_2^{(1)},\dots)\,,\end{equation}
and the mass matrix is 
\begin{equation}\label{MassMatrix}
\mathcal{M}=
\begin{pmatrix}
0 & \hat{m}_{1}^{(0)} & \hat{m}_{2}^{(0)} &  \hat{m}_{1}^{(1)} & \hat{m}_{2}^{(1)} & \dots\\
 \hat{m}_{1}^{(0)} & m_1^{(0)} & 0 & 0 & 0& \dots\\
\hat{m}_{2}^{(0)} &0 &m_2^{(0)} &0 & 0& \dots\\
\hat{m}_{1}^{(1)} &0 & 0&m_1^{(1)} &0 & \dots\\
\hat{m}_{2}^{(1)} &0 & 0& 0&m_2^{(1)} & \dots\\
\vdots &\vdots & \vdots&\vdots &\vdots & \ddots\\
\end{pmatrix}\,.
\end{equation}
Given the cutoff scale $\Lambda$, above which the theory becomes non-perturbative, it is possible to determine how many particles will contribute to the mass matrix. It is  usually assumed $\Lambda L =20$ for UED models \cite{Beuria:2017jez}, thus in our case, considering $L^{-1}=2$ TeV to avoid LHC constraints, we have $\Lambda=40$ TeV. KK particles heavier than the cutoff scale are not present in the mass matrix, thus there are roughly $\Lambda R$ roots (sterile neutrino KK particles) below the cutoff scale.

The  characteristic eigenvalue equation $\det(\mathcal{M}-I\lambda)=0$ that determines the physical neutrino masses is written for the mass matrix (\ref{MassMatrix}) as
\begin{equation}\label{eq:eigenvalueeq}
       \prod_n(m_1^{(n)}-\lambda)(m_2^{(n)}-\lambda)\biggr[\lambda + \sum_n\biggr(\frac{\Hat{m}_1^{(n)\,2}}{m_1^{(n)}-\lambda}+\frac{\Hat{m}_2^{(n)\,2}}{m_2^{(n)}-\lambda}\biggr)\biggr]=0\,.
\end{equation}
As   can seen from the mass matrix, the smaller the couplings $\Hat{m}_{1(2)}^{(n)}$ are, the smaller the KK masses need to be to satisfy the neutrino mass $m_{\nu_{L}}\sim 10^{-2} $ eV. If a very large number of particles enter in the mass matrix, the seesaw mechanism may not be achieved because there would be contributions of a large amount  of modes, increasing the smallest mass (SM neutrino mass). Since $\Lambda=40$ TeV for $L^{-1}=2$ TeV,  the number of sterile neutrino KK states below the cutoff scale are $\Lambda R\sim 4\times 10^4$, for $R^{-1}=1$ GeV, being this number larger for larger radii. Such a large number of KK particles could leave the SM neutrino heavier than it should be, unless  very small 4-D Yukawa couplings $\lambda_{4, 1(2)}$ are assumed.  On the other hand, if the bulk Dirac mass is of the same order of magnitude of  the cutoff scale $\Lambda$ ($\mathcal{O}(10 $) \text{TeV}), in such a way that a relatively small number of particles contribute to the mass matrix, this issue can be easily avoided and the seesaw mechanism can be properly achieved. 

Since the sterile neutrinos are much heavier than the off-diagonal masses $m_{1(2)}^{(n)}\gg\Hat{m}_{1(2)}^{(n)}$, Eq. (\ref{eq:eigenvalueeq}) can be simplified because all of the eigenvalues but the correspondent one to the SM neutrino mass are practically equal to the respective masses, that is, $m_{1(2)}^{(n)}\sim \lambda_n$. For the SM neutrino, the corresponding eigenvalue $\lambda_\nu$ is obtained from the term in square brackets in Eq. (\ref{eq:eigenvalueeq}), which gives the following result after making the approximation $m_{1(2)}^{(n)}\gg \lambda_\nu $,
\begin{equation}\label{eq:eigenvalueeqNeutrino}
     \lambda_\nu \approx - \sum_n\biggr(\frac{\Hat{m}_1^{(n)\,2}}{m_1^{(n)}}+\frac{\Hat{m}_2^{(n)\,2}}{m_2^{(n)}}\biggr)\,.
\end{equation}
We see that the standard seesaw expression is obtained if there is only one particle ($\lambda_\nu\approx -\Hat{m}_1^2/m_1$). From Eq. (\ref{eq:eigenvalueeqNeutrino}) we also understand why a very large number of particles would increase $\lambda_\nu$, jeopardizing the success of the seesaw mechanism.

It is possible to have an estimate for the upper limit of $\hat{m}^{(n)}_{1(2)}$ that would give the observed neutrino mass $\lambda_\nu \sim 10^{-2}$ eV. We may consider $m_M\sim0$ for a moment for simplicity. When the Majorana mass in the bulk is absent, the two sterile neutrino towers  have degenerate mass states $m_1^{(n)}= m_2^{(n)}$. For large  values of the  bulk Dirac mass, the physical masses are roughly the same for almost all KK states. Additionally, just to have an intuition for the values of $\hat{m}_{1(2)}^{(n)}$, let us assume that $\hat{m}_{1}^{(n)}=\hat{m}_{2}^{(n)}\equiv \hat{m}_1$, \textit{i.e.}, it is independent of $n$. This is not completely true but gives a conservative estimate, because $\hat{m}_{1(2)}^{(n)}$ is smaller for small or very large $n$.  Therefore, in this situation Eq. (\ref{eq:eigenvalueeqNeutrino}) yields
\begin{equation}
     \Hat{m}_1^2\lesssim 10^{-11}  \frac{m_1}{ \Lambda R} \text{GeV}\,,
\end{equation}
where the sum over $n$ becomes the product of $\Lambda R$ states. For $R=1$ GeV, the number of states are  $\Lambda R\sim 4\times 10^4$, therefore, for $m_1 \sim 10$ TeV we should have $ \Hat{m}_1\lesssim 10^{-5}$ GeV. Obviously this is just a rough estimate, but it gives an idea of how easy the seesaw mechanism could be satisfied within the present model, having the lightest sterile neutrinos with masses of order $10\;\text{TeV}$. On the other hand, larger compactification radii would lead to a much larger number $\Lambda R$ of states, therefore requiring smaller couplings for the same neutrino mass.

 We numerically solve Eq. (\ref{eq:eigenvalueeq})   and gather in Table \ref{tab:parameters} some representative and  plausible values of parameters that satisfy what is expected for the seesaw mechanism $m_\nu\sim 10^{-2}$ eV.  Other choices of parameters would give similar results. For $R^{-1}=100 $ MeV there are over $10^5$ KK states that contribute to the neutrino mass matrix (considering the values of bulk masses in Table \ref{tab:parameters}), while for $R^{-1}=1$ GeV there are  $\sim 10^4$ states. Although the couplings are relatively more suppressed for larger compactification radii, they are not sufficiently small to compensate the additional number of KK states contributing to the neutrino mass in Eq. (\ref{eq:eigenvalueeqNeutrino}). Therefore, in order to compensate the eventually large number of KK states for larger $R$, smaller 4-D Yukawa couplings $\lambda_{4,1(2)}$ are needed. We see from Table \ref{tab:parameters} that the most favorable compactification radius is $R^{-1}=1$ GeV. Smaller values of the 4-D Yukawa couplings, $\lambda_{4,1(2)}\sim 0.01-0.1$ for $R^{-1}=1$ GeV, for instance would bring the upper limits in  Table \ref{tab:parameters} down to smaller values, $m_D\sim 1-10$ TeV, respectively. The same reasoning also applies to obtain smaller values for the BLKT parameter $\delta_A \alpha$.  Notice that the roots $x_n$ are usually much smaller than $m_D$, leading to a difference between one KK state and the next one of order $(x_{n+1}^2-x_{n}^2)/(2m_D R^2)$. The mass difference between two neighboring KK states can be much smaller than the difference between the masses of the two sterile neutrinos within the same KK state, that is, $m_{1(2)}^{(n)}-m_{1(2)}^{(k)}\sim (x^2_n-x^2_k)/(2m_D R^2)\ll m_1^{(n)}-m_2^{(n)}\sim m_M$, provided that $m_M$ is not very small.
\begin{table}[h]
    \centering
    \begin{tabular}{|c|c|c|c|c|}
    \hline
      $R^{-1} $  & $m_D$  &$m_M$  & $\delta_A \alpha$  &$\lambda_{4,1(2)}$\\
      \hline
       1 GeV  &$\geq 30$ TeV &$\leq 10$ TeV & $\geq 30$ TeV & $ \leq 1$\\
       100 MeV  &$\geq 30$ TeV &$\leq 10$ TeV & $\geq 30$ TeV& $\leq 0.1$ \\
        10 MeV  &$\geq 30$ TeV &$\leq 10$ TeV & $\geq 30 $  TeV& $\leq 0.01$ \\
         \hline
             \end{tabular}
    \caption{Representative set of parameters, and conservative upper or lower limits that satisfy the SM neutrino eigenvalue $m_{\nu_L}\sim 10^{-2}$ eV, for $L^{-1}=2$ TeV,  and $R^{-1}=10^{-2}, 10^{-1},1$ GeV. 
    }
    \label{tab:parameters}
\end{table}

\begin{figure}
    \centering
    \includegraphics[scale=0.5]{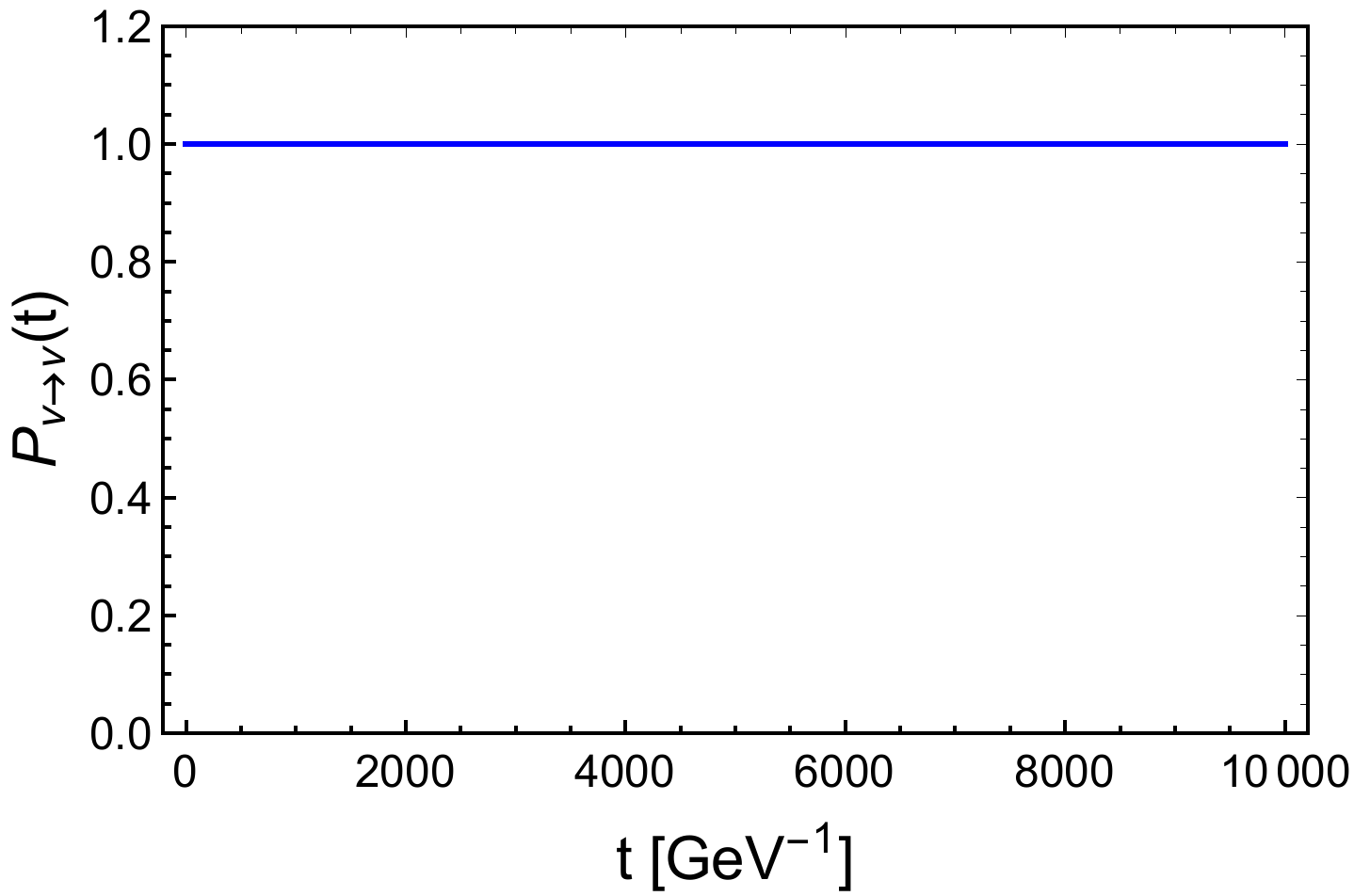}
    \caption{Survival probability as a function of time for the SM neutrino, using the last row of values in Table \ref{tab:parameters}. Different values of the parameters give quite similar results. }
    \label{fig:survivalprob}
\end{figure}

Finally, in order to understand the influence of the KK particles on neutrino oscillations we will evaluate the total probability of the SM neutrino oscillating into any other sterile neutrino KK state. It is convenient to work with the survival probability $P_{\nu_{L}\rightarrow \nu_{L}}(t)$,  as a function of time, that the SM neutrino is preserved \cite{Dienes:1998sb}
\begin{equation}\label{eq:susvivalprob}
  P_{\nu_{L}\rightarrow \nu_{L}}(t)  =\Big| \sum_i |U_{\nu_{L}i}|^2\exp(i E_i t)\Big|^2\,,
\end{equation}
where the energies $E_i$ are the mass eigenvalues in our case and $U_{\nu_{L}i}$ are the  mass eigenvectors. The gauge eigenstates are therefore written in terms of the mass eigenstates as
\begin{equation}
    \mathcal{N}=U \mathcal{\Tilde{N}}\,,
\end{equation}
where the mass eigenvectors $U$ are
\begin{equation}
U=    \begin{pmatrix}
U_{\nu}\\
U_{1,0}\\
U_{2,0}\\
U_{1,1}\\
U_{2,1}\\
\vdots
\end{pmatrix}\,.
\end{equation}
Each row is the eigenvector correspondent to the eigenvalue   $E_i=\lambda_i$, and it can be written as
\begin{equation}\label{eq:eigenvectors}
    U_i=\begin{pmatrix}
    1,& \frac{\hat{m}_1^{(0)}}{m_1^{(0)}-\lambda_i},&  \frac{\hat{m}_2^{(0)}}{m_2^{(0)}-\lambda_i},& \frac{\hat{m}_1^{(1)}}{m_1^{(1)}-\lambda_i} ,&  \frac{\hat{m}_2^{(1)}}{m_2^{(1)}-\lambda_i},& \dots,&  \frac{\hat{m}_1^{(k)}}{m_1^{(k)}-\lambda_i}, &\frac{\hat{m}_2^{(k)}}{m_2^{(k)}-\lambda_i},&\dots
    \end{pmatrix}\,,
\end{equation}
where $i \neq k$. Using the parameters  in Table \ref{tab:parameters} it is possible to check that all the terms $\frac{\hat{m}_{1(2)}^{(k)}}{m_{1(2)}^{(k)}-\lambda_i}$ are very small. Therefore, the survival probability (\ref{eq:susvivalprob}) remains very close to one, as can be seen in Fig. \ref{fig:survivalprob} for some specific values of parameters, although for other values the results are qualitatively the same.

Due to the large sterile neutrino masses, experimental/observational constraints do not pose challenges to this model \cite{Atre:2009rg,Boyarsky:2009ix,Drewes:2013gca,Drewes:2015iva,Bolton:2019pcu}. Furthermore, an interaction  between the sterile neutrinos with the weak gauge bosons would arise from $\nu_{L \alpha}=U_{\alpha i} \nu_i+\Theta_{\alpha 1(2)}^{(n)}N_{1(2)}^{(n)}$, where $\nu_i$ $(i=1,2,3)$ are the active neutrinos and $\Theta_{\alpha 1(2)}^{(n)}\equiv  \hat{m}_{1(2)}^{(n)}/m_{1(2)}^{(n)}$  comes from Eq.  (\ref{eq:eigenvectors}). This  admixture of the tower of sterile neutrinos with SM neutrinos, $\Theta_{\alpha 1(2)}^{(n)}$, agrees with the case where there are only two sterile neutrinos \cite{Asaka:2011pb}. For the values presented in Table \ref{tab:parameters} one can get $\Theta_{\alpha 1(2)}^{(n)}\leq 10^{-9}$ (and even smaller values for slightly smaller 4-D Yukawa couplings $\lambda_{4, 1(2)}$). Therefore the mixing of sterile neutrinos with active SM neutrinos is extremely small, not modifying significantly the weak currents. 

\section{Leptogenesis}\label{sec:leptogenesis}
In this section we investigate whether baryogenesis via leptogenesis is viable in the present model.
Since the two KK towers imply that there is a large number of sterile neutrinos contributing to the seesaw mechanism,
their individual couplings to Higgs bosons are small compared to the usual scenarios with no more than three
sterile neutrinos. We recall from Section~\ref{sec:seesaw} that   the number of KK sterile neutrinos is limited by the cutoff scale such that, in order to avoid contributions to the masses of active neutrinos from a very large number of sterile states, we choose $m_D\lesssim m_{1,2}^{(n)}\lesssim\Lambda$, e.g. $m_D=30\,{\rm TeV}$ and $\Lambda =40\,{\rm TeV}$ in our parametric examples. 

\begin{figure}
    \centering
    \includegraphics[scale=0.6]{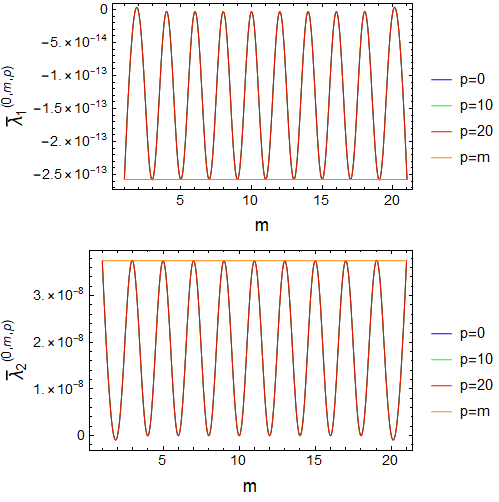}
    \caption{Yukawa couplings (of the lightest sterile neutrino) as a function of the KK SM neutrino states ($m)$, for different KK  Higgs states ($p$), for $m_D=30$ TeV, $m_M=10^{-9}$ GeV, $\lambda_{4,1(2)}=10^{-2}$, $R^{-1}=1$ GeV, $L^{-1}=2$ TeV and $\delta_A\alpha=10$ TeV. The couplings have practically the same order of magnitude as for the SM zeroth-mode and the results are the same for different sterile neutrino states. }
    \label{fig:plotKKSM}
\end{figure}

For this setup, notice that at high temperatures $T\sim m_1^{(0)}$, it is expected that the SM KK states also contribute as final states in the sterile neutrino decay. As explained previously, the number of SM KK states is $\Lambda L$ (where we take $\Lambda L=20$ for our parametric examples). The sterile neutrino decay rate is therefore the sum over the rates for all final states $N^{(n)}_{1,2}\rightarrow h^{(m)} +\nu^{(p)}$. The individual Yukawa couplings are no longer described by Eq. (\ref{eq:couplings}) because they should include, in the integral over the ED, the contribution of the excited KK states of the Higgs and the SM neutrino. Their corresponding wave functions are \cite{Appelquist:2000nn}
\begin{equation}
    \nu^{(m>0)}(h^{(p>0)})(y)=\sum_{m(p)=1}^{\Lambda L}\frac{\sqrt{2}}{\sqrt{\pi L}}\cos\left(\frac{m(p) }{L}y\right)\,,
\end{equation}
and the effective Yukawa couplings for $m,n \neq 0$ become  
\begin{equation}
    \Bar{\lambda}_{1(2)}^{(n,m,p)}\equiv \frac{2\lambda_{4,1(2)}}{\Lambda_0\pi L} \int_{\pi r}^{\pi R} \,dy f_{1(2)\text{II}}^{(n)}(y)\cos\left(\frac{m }{L}y\right)\cos\left(\frac{p }{L}y\right)\,.
\end{equation}
It turns out, for  the range of values of the parameters to be considered here (as presented below), the Yukawa couplings have practically the same order of magnitude over all of the KK spectrum, as depicted in Fig. \ref{fig:plotKKSM}.

\begin{figure}
    \centering
    \includegraphics[scale=0.6]{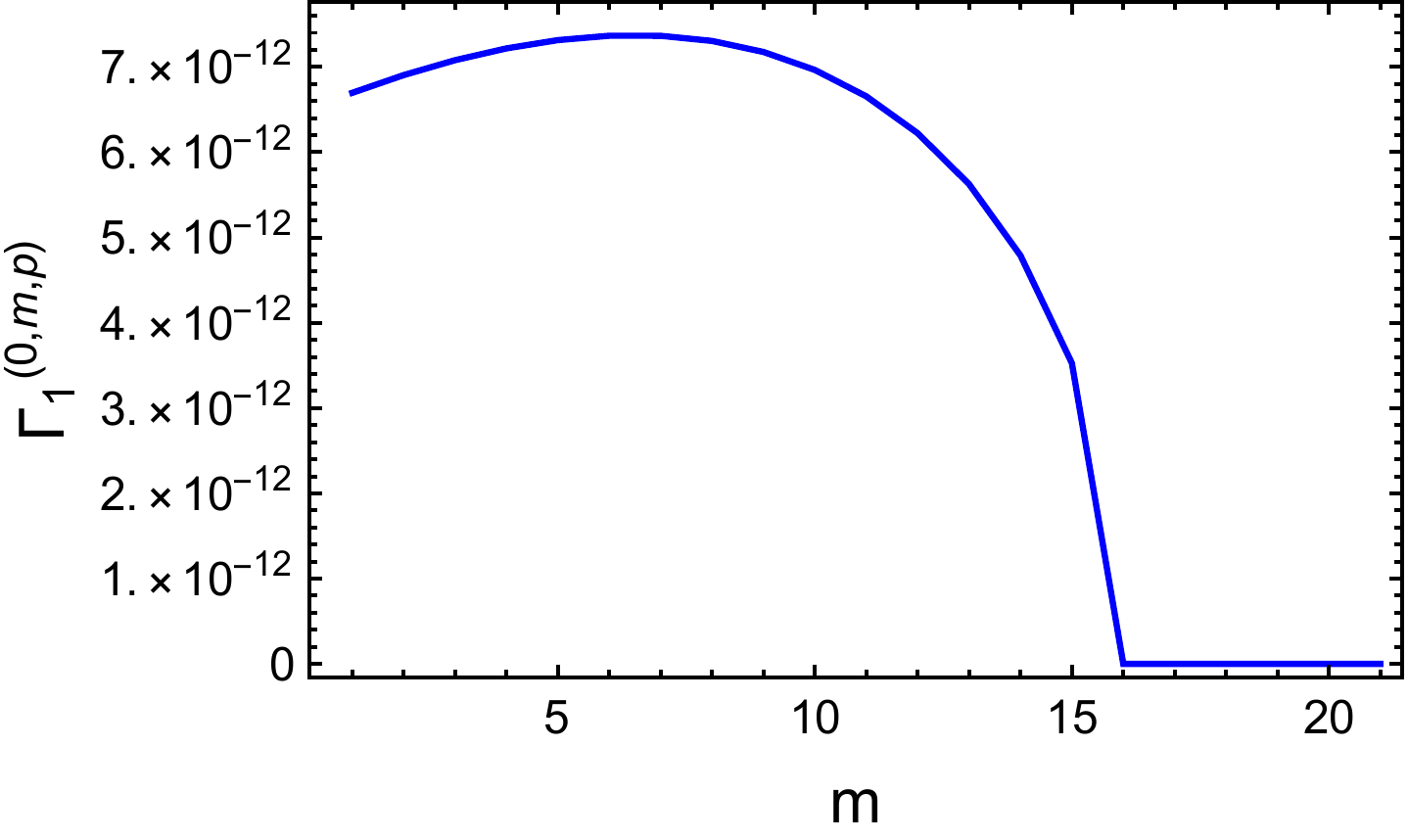}
    \caption{Sterile neutrino decay rate (of the lightest  particle $N_1^{(0)}$) into  KK Higgs and neutrino, as a function of the KK SM neutrino states ($p=1$ fixed, although it gives similar  results for different  Higgs states) for $m_D=30$ TeV, $m_M=10^{-9}$ GeV, $\lambda_{4,1(2)}=10^{-2}$, $R^{-1}=1$ GeV, $L^{-1}=2$ TeV and $\delta_A\alpha=10$ TeV. The decay rate is roughly the same of the SM zeroth-mode $(m=0)$ and  the results are analogous for different sterile neutrino states. 
    }
    \label{fig:decayKK}
\end{figure}

In UED models the mass spectrum of the SM particles is given by $m_{SM}^{(p)}=\sqrt{m_{SM}^2+p^2/L^2}$, where $p=0,1,2, \dots$ and $m_{SM}$ is the mass of the known SM particles. As it has been discussed before, for our choice of parameters, there are $\Lambda L=20$ roots below the cutoff scale, and since $L^{-1}=2$ TeV, the excited SM KK states have masses  of approximately $p/L(\gg m_{\text{SM}})$. The decay rate of the sterile neutrinos into SM KK states is of the same order of magnitude as for the corresponding zeroth SM-mode that is shown in Fig.~\ref{fig:decayKK}.

Therefore, for the sake of simplicity and without loss of generality, in the following we can use the expressions for the sterile neutrino decay into only SM zero-modes, but to take into account the final SM KK states as well, we  multiply the results of the SM zero-mode by the number of kinematically allowed processes $\sim \Lambda L(\Lambda L+1)/2$. This estimate holds by order of magnitude since we assume that
$m_D\lesssim m_{1,2}^{(n)}\lesssim\Lambda$.

The parameter characteristic for leptogenesis is the washout strength
\begin{align}
K_{1,2}^{(n)}=\frac{\Gamma(N_{1,2}^{(n)}\to L^{(0)}H^{(0)})}{H|_{T=m_{1,2}^{(0)}}}=\frac{|\lambda_{1,2}^{(n)}|^2 m_{1,2}^{(n)} M_{\rm Pl}}{32\pi \sqrt{g_*} m_{1,2}^{(0)2}}\,,
\end{align}
where the subscripts and superscript refer to the sterile neutrino $N_{1,2}^{(n)}$, $\Gamma$ is the decay rate in vacuum,
$H$ is the Hubble rate, $T$ is the temperature, $g_*$ is the number of degrees of freedom for relativistic particles, and $M_{\rm Pl}$ is the Planck mass. Due to the couplings being
weak here in comparison with the usual seesaw scenarios, $\Lambda L(\Lambda L+1)/2\;K_{1,2}^{(n)}\ll1$ for typical configurations in parameter space. This
relation implies weak washout, that is, the sterile neutrinos $N_{1,2}^{(n)}$ remain far from equilibrium before their
distribution becomes Maxwell-suppressed, and each individual sterile neutrino only washes out a small fraction of the lepton asymmetry. Note that we
assume here that the initial abundances of the sterile neutrinos vanish.

\begin{figure}
    \centering
    \includegraphics[scale=0.4]{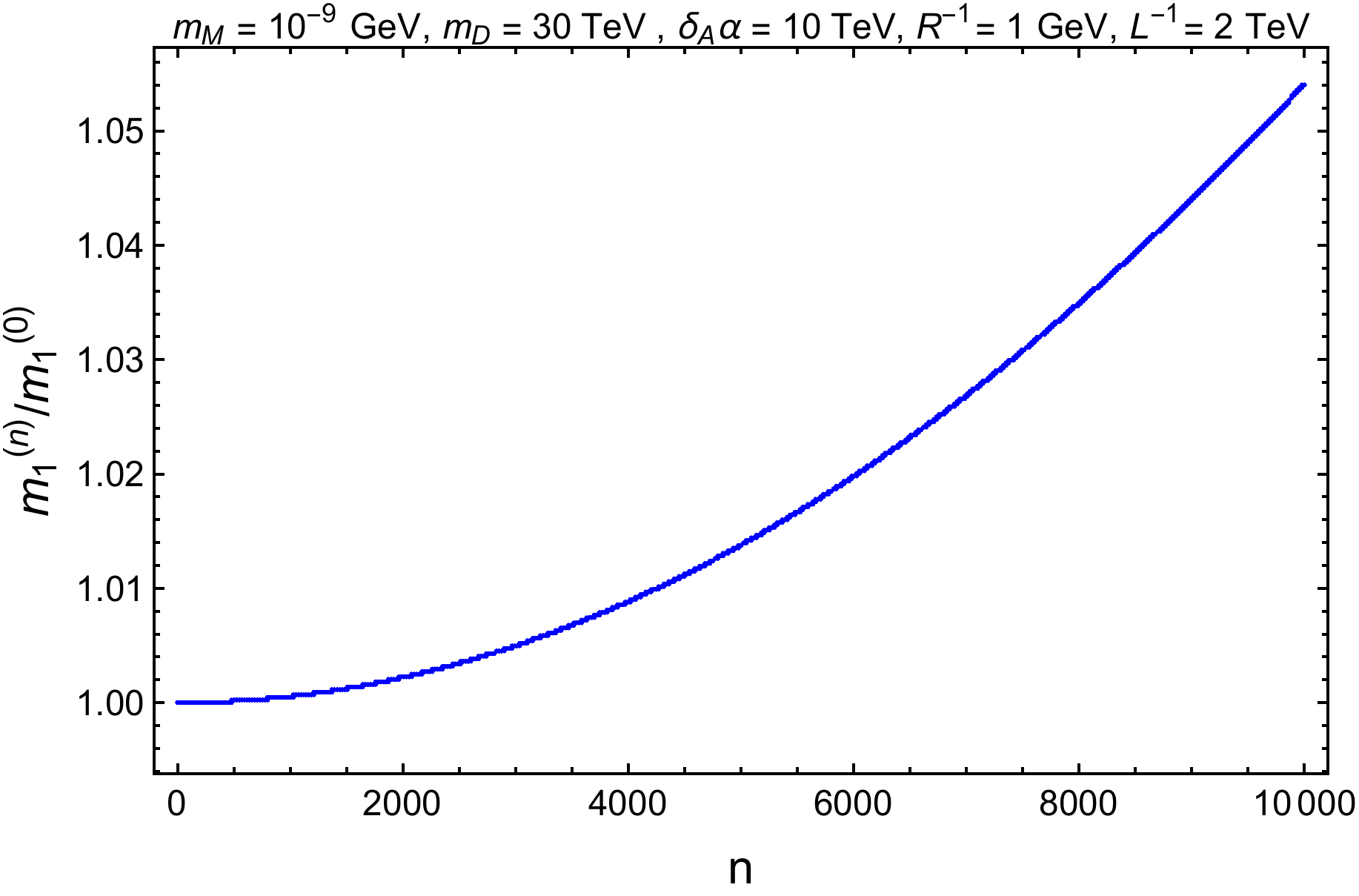}
    \caption{Masses of the sterile neutrino KK states normalized to the mass of the lightest state, for  for $m_D=30$ TeV, $m_M=10^{-9}$ GeV, $R^{-1}=1$ GeV, $L^{-1}=2$ TeV and $\delta_A\alpha=10$ TeV. }
    \label{fig:massrate}
\end{figure}

Some studies of leptogenesis with a large number of sterile neutrinos~\cite{Eisele:2007ws} were carried out before the relevant
reaction rates for sterile neutrinos in the relativistic regimes were thoroughly investigated~\cite{Besak:2012qm,Garbrecht:2013gd,Ghisoiu:2014ena}. As a consequence,
the dynamics of leptogenesis in scenarios with many sterile neutrinos should be reconsidered in detail, which
is beyond the scope of the present work. To obtain an estimate of the asymmetry, we rely on the work~\cite{Garbrecht:2019zaa}.
Its main shortcoming when applied to the present scenario is the assumption of a hierarchical spectrum of
sterile neutrinos which does not apply to the present setup where the main contributions to the asymmetry arise from the resonant mixing
of pairs of sterile neutrinos $N^{(n)}_{1}$ and $N^{(n)}_{2}$ that are close in mass.
While the resonant enhancement can be included in the appropriate factor describing the
decay asymmetry this leaves an inaccuracy in the efficiency factor of order one.
We show in Fig. \ref{fig:massrate} the mass hierarchy between the sterile neutrino KK excited states and its zero-mode, for one sterile neutrino and some representative parameter choices, although other values give similar behavior.
 Nonetheless, we can make a prediction of order one accuracy
when considering only the lightest $2\bar n$ of the sterile neutrinos such that these will not wash out of most of the
produced asymmetry. That is, we set $\bar n$ by the condition
\begin{align}
\frac{\Lambda L(\Lambda L+1)}{2} \sum\limits_{n=0}^{\bar n}\left(K_1^{(n)}+K_2^{(n)}\right)\geq 1\,,
\end{align}
which is to be understood as an estimate. 

Returning to our original discussion, for each individual sterile neutrino, the decay asymmetry is given by
\begin{align}
    \varepsilon^{(n)}_{1,2}&=\frac{\Lambda L(\Lambda L+1)}{2}\frac{\Gamma(N_{1,2}^{(n)}\rightarrow L^{(0)}H^{(0)})-\Gamma(N_{1,2}^{(n)}\rightarrow \bar{L}^{(0)}\bar{H}^{(0)})}{\Gamma(N_{1,2}^{(n)}\rightarrow L^{(0)}H^{(0)})+\Gamma(N_{1,2}^{(n)}\rightarrow \bar{L}^{(0)}\bar{H}^{(0)})}\nonumber\\
    & \approx \frac{\Lambda L(\Lambda L+1)}{2}\sum\limits_{k=0}^{\bar n}\frac{\text{Im}\big[(\lambda_{1,2}^{(n)* }\lambda_{2,1}^{(k)})^2\big] }{8\pi |\lambda_{1,2}^{(n)}|^2}\biggr[f\left(\frac{m_{2,1}^{(k)\,2}}{m_{1,2}^{(n)\,2}}\right)+g\left(\frac{m_{2,1}^{(k)\,2}}{m_{1,2}^{(n)\,2}}\right)\biggr]\,,
   \end{align}
   where
\begin{align}
    f(x)&=\sqrt{x}\left[1-(1+x)\ln\left(\frac{1+x}{x}\right)\right]\,,\\
    g(x)&\approx\frac{\sqrt{x}}{1-x}\,,
    \end{align}
and where we have again included the factor accounting for the enhancement due to the SM KK states.
The expression for $g(x)$ is valid in the limit $|m_{1(2)}^{(k)}-m_{2(1)}^{(n)}|\gg |\Gamma_{1(2)}^{(k)}-\Gamma_{2(1)}^{(n)}|$, which is the case for this model.  The phase of the 4-D Yukawa couplings $\lambda_{1}^{(n)}=(e^{i\phi_1^n}\Bar{\lambda}_1^{(n)}+e^{i\phi_2^n}\Bar{\lambda}_2^{(n)})/\sqrt{2}$ and $\lambda_{2}^{(n)}=-i(e^{i\phi_1^n}\Bar{\lambda}_1^{(n)}-e^{i\phi_2^n}\Bar{\lambda}_2^{(n)})/\sqrt{2}$ \cite{Pilaftsis:1999jk} is not relevant in the following discussion, as long as $\text{Im}\big[(\lambda_1^{(n)* }\lambda_2^{(k)})^2\big] \sim \sin(\phi_1^n-\phi_2^n)\neq 0$.

Provided we can neglect the washout by the above assumptions, the efficiency factor in the weak washout regime is given by $\kappa_{1,2}^{(n)}=\frac{\Lambda L(\Lambda L+1)}{2}0.32 K_{1,2}^{(n)}$~\cite{Garbrecht:2019zaa}.
Note that the weak washout approximation can only be applied provided $\frac{\Lambda L(\Lambda L+1)}{2}0.32 K_{1,2}^{(n)}\ll 1$.
The contribution of $N_{1,2}^{(n)}$ to the final asymmetry
then is $Y_{B-L;1,2}^{(n)}=- \varepsilon^{(n)}_{1,2} \kappa_{1,2}^{(n)} Y_{N{\rm eq}}(t=0)$, where $Y_{N{\rm eq}}(t=0)$
is the yield of a sterile neutrino in the relativistic regime. Summing over all sterile neutrinos up to $N_{1,2}^{(\bar n)}$, we arrive at
\begin{align}
\label{BAU:final}
Y_{B-L}=-\sum\limits_{n=0}^{\bar n} \sum\limits_{i=1}^2 \varepsilon^{(n)}_{i}  \frac{\Lambda L(\Lambda L+1)}{2} 0.32 K_{i}^{(n)} \frac{T^3}{s}\,,
\end{align}
where $s$ is the entropy density $s\sim g_* T^3$.


The fraction involving the Yukawa couplings in $\varepsilon^{(n)}_{1,2}$ is very small, so that in order to compensate it, the function $g(x)$ should be large enough to eventually give the observed baryon asymmetry \cite{Chen:2007fv}. The function $g(x)$ is  large for $m_1^{(n)} \sim m_2^{(n)}$. However, since the sums are over $k$ and $n$, there is a large contribution from the KK tower of states, and the particular combination of $g(x)$ with the Yukawa couplings such as to attain the observed asymmetry requires a specific choice of parameters. 
    
\begin{figure}
    \centering
    \includegraphics[scale=0.5]{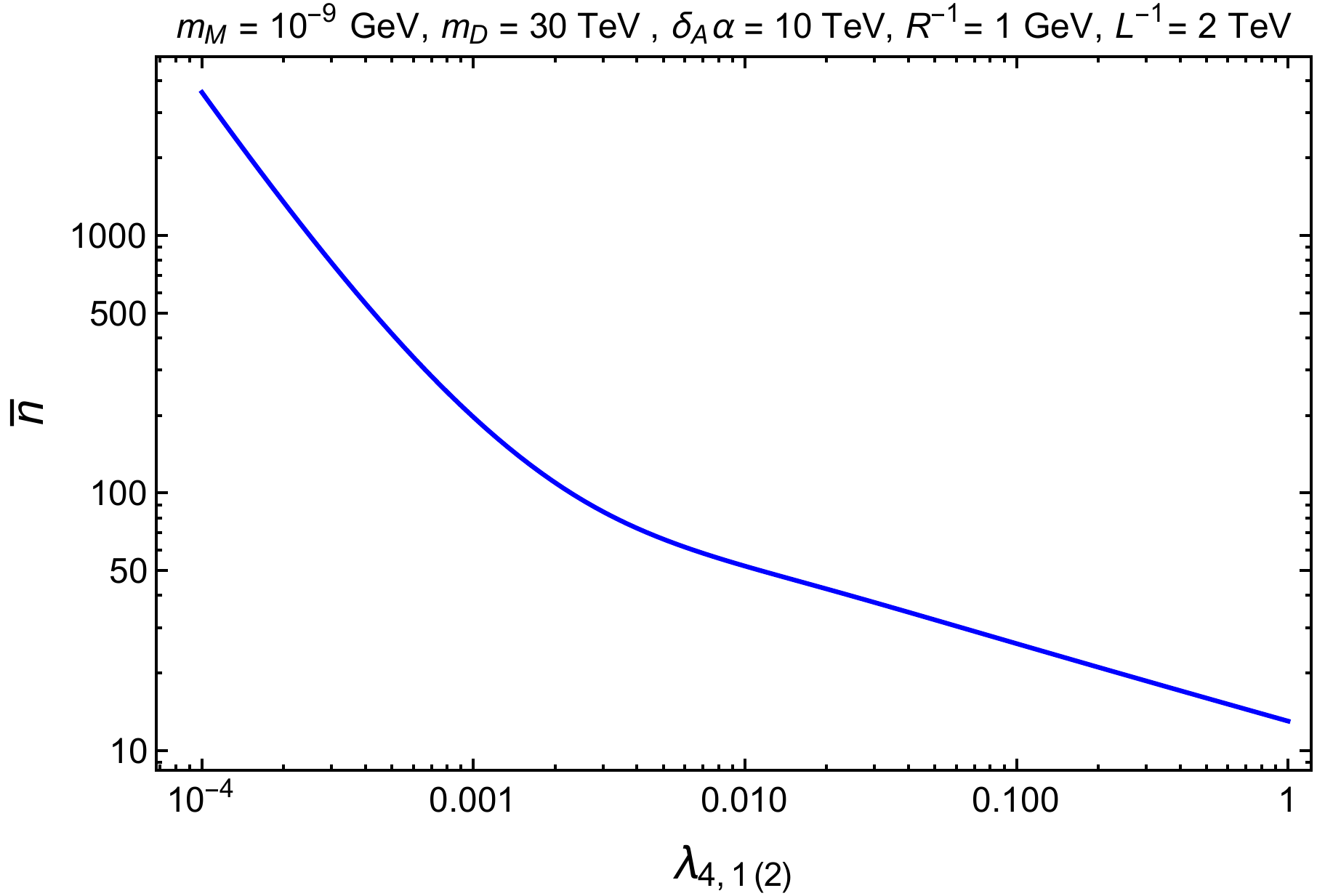}
    \caption{Maximum number of KK sterile neutrino states $\Bar{n}$ that are not washed out, as a function of the 4-D Yukawa coupling $\lambda_{4,1}=\lambda_{4,2}$. }
    \label{fig:washoutNbar}
\end{figure}

In order to illustrate the parametric dependence of the asymmetry, we let one parameter free, while the other ones are kept fixed, considering always $R^{-1}=1$ GeV and $L^{-1}=2$ TeV for simplicity, although other values give similar results. We evaluate the maximum number of KK sterile neutrino states $\Bar{n}$ which are not washed out, for a set of parameters, as a function of the 4-D Yukawa couplings and show this in Fig. \ref{fig:washoutNbar}. We take $\lambda_{4,1}=\lambda_{4,2}$, as $\lambda_{4,1}\neq \lambda_{4,2}$ gives similar and interpolating results for $\bar n$. 

\begin{figure}
    \centering
    \includegraphics[scale=0.5]{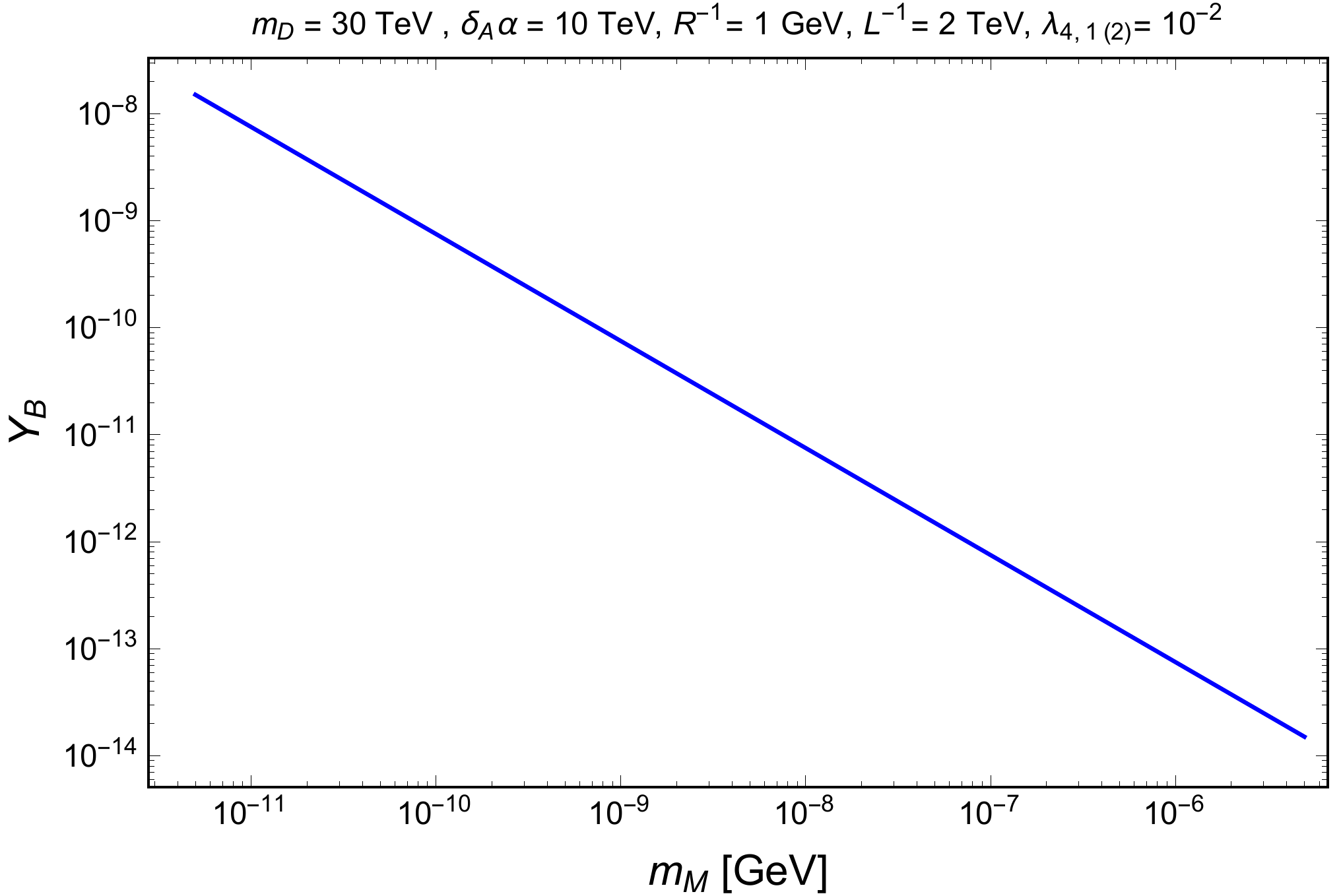}
    \caption{Baryon asymmetry $Y_B$ as a function of the bulk Majorana mass $m_M$, for a specific set of parameters. The observed value of the baryon asymmetry is obtained for $m_M\sim 8 \times 10^{-10}$ GeV. 
    }
    \label{fig:Mmajlepto}
\end{figure}

\begin{figure}
    \centering
    \includegraphics[scale=0.5]{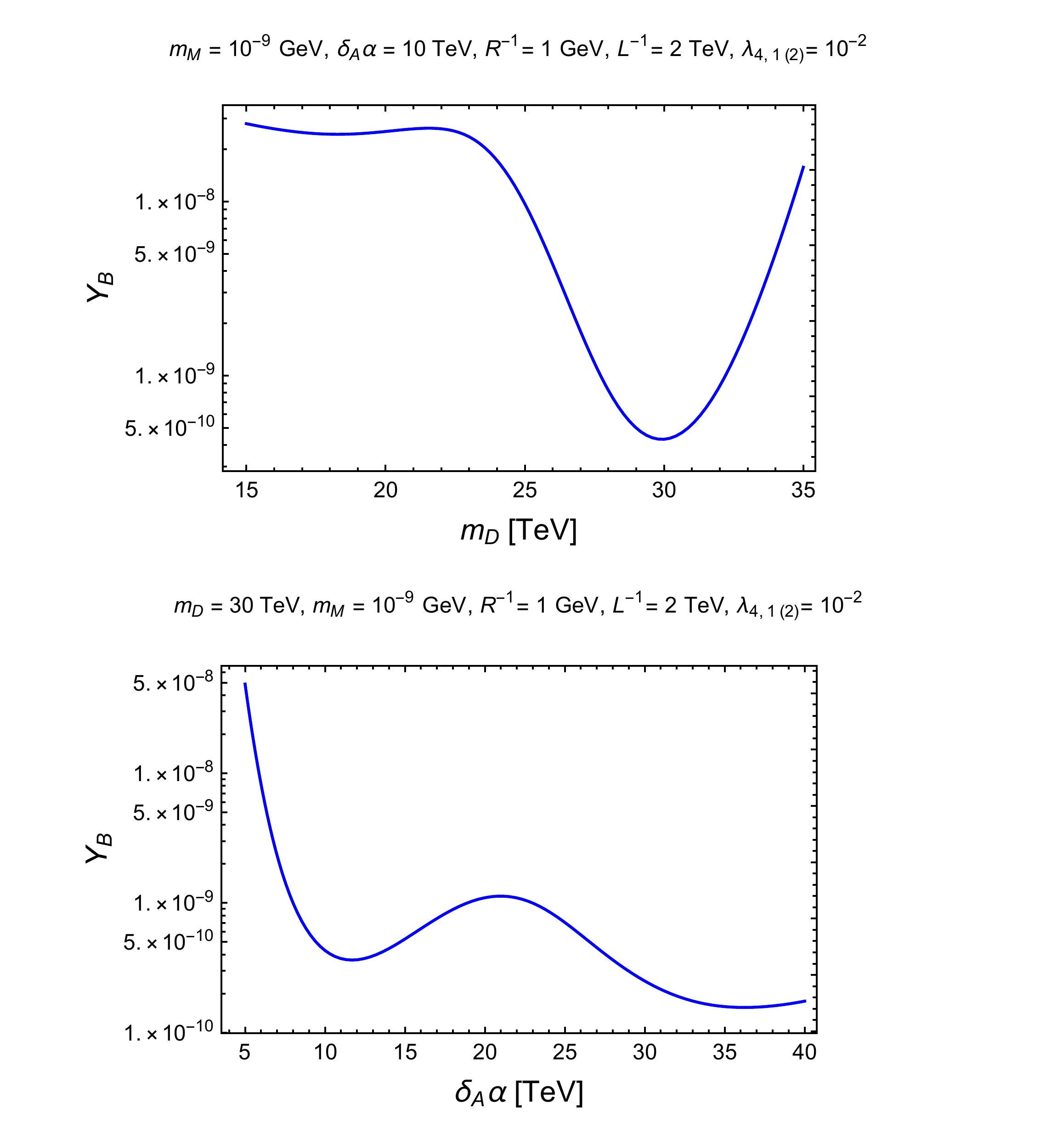}
    \caption{Baryon asymmetry $Y_B$ as a function of the bulk Dirac mass (top) and the BLKT parameter $\delta_A\alpha$ (bottom), while the other parameters are fixed. The baryon asymmetry is only obtained for $m_D\sim 30$ TeV.}
    \label{fig:MDlepto}
\end{figure}

\begin{figure}
    \centering
    \includegraphics[scale=0.5]{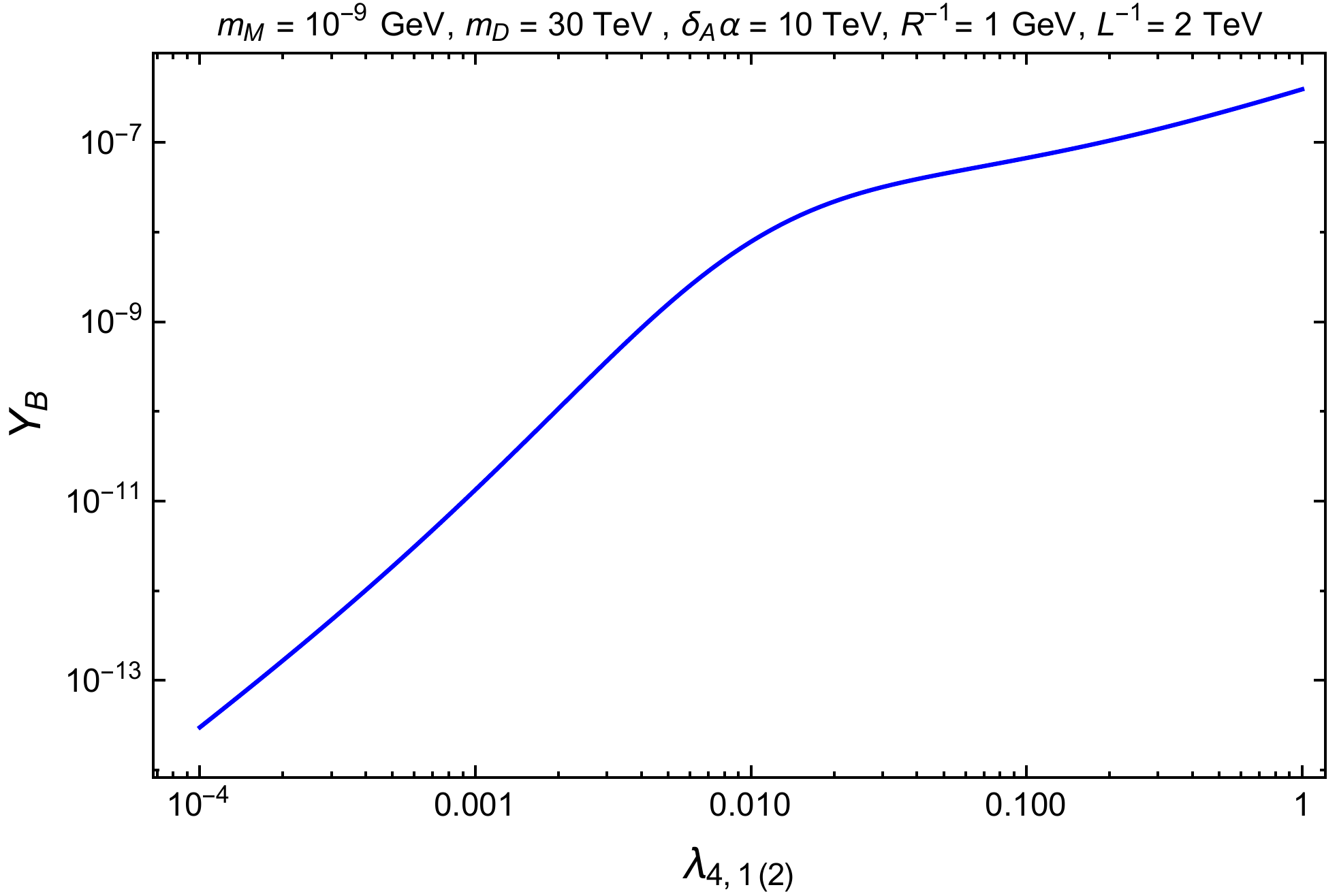}
    \caption{Baryon asymmetry as a function of the 4-D Yukawa couplings $\lambda_{4,1}=\lambda_{4,2}$. }
    \label{fig:plotlambda}
\end{figure}

Figs. \ref{fig:Mmajlepto}, \ref{fig:MDlepto} and \ref{fig:plotlambda} show the baryon asymmetry as a function of one free parameter. It can be seen that, in order to explain the baryon asymmetry, the bulk masses should be $m_D\sim 30$ TeV and $m_M\sim 10^{-9}$ GeV, while the BLKT parameter  can have various values. Finally, in Fig. \ref{fig:plotlambda} the baryon asymmetry is shown as a function of the 4-D Yukawa couplings. Although we set $\lambda_{4,1}=\lambda_{4,2}$, for simplicity, other values of these parameters would give essentially similar results. While the resonant enhancement of the asymmetry can be naturally achieved for small values of $m_M$, we note that the model does not predict a preferred value for the baryon asymmetry. Rather, the freedom of choice for the parameters
leads to a wide range of predictions around the observed value.

The result~(\ref{BAU:final}) can be compared with the well-known realizations of resonant leptogenesis in the absence of ED, where one should of course set $\bar n=0$. More importantly, while in the present case $K_i^{(n)}\ll 1$ given the light neutrino masses and the large number $\Lambda L$ of KK
states contributing to the seesaw mechanism, in the absence of ED the washout strength generically satisfies $K_i^{(0)}\gtrsim 30$, barring
the decoupling of individual sterile neutrinos~\cite{Pilaftsis:2003gt}. (We set $n=0$ because in this case, there are no KK modes. In addition, one typically assumes a small number, i.e. two or three sterile neutrinos). Hence, as stated above, in the ED scenario with BLKT, we are in the weak washout regime of leptogenesis whereas without ED, resonant leptogenesis typically takes place in the strong washout regime. In that case, the factor $\sim K_i^{(n)}$ in Eq.~(\ref{BAU:final}) gets replaced with $\sim 1/K_i^{(0)}$, where the order one coefficients and higher order corrections are discussed in the literature~\cite{Garbrecht:2019zaa,Chen:2007fv}. Both the dynamics and the parametric outcome in the present scenario of leptogenesis are therefore different when comparing with the case without ED.

\section{Conclusions}\label{sec:conclu}
In this paper we have shown that the seesaw mechanism can occur in an ED scenario for sterile neutrinos as light as 1--10 TeV, for instance. We have considered a flat and single ED with a fat brane at one end of the interval, where the SM is confined, thus having a spectrum similar to UED models in 5-D. Only a Dirac fermion is present in the bulk and due to the BLKT the interaction between the resulting two towers of 4-D  Majorana sterile neutrinos and the Higgs can be very suppressed. Thus, it is not
that the sterile neutrinos are very massive in order to explain the SM neutrino mass. We have presented illustrative calculations using compactification radii of $R^{-1}=10^{-2}, 10^{-1}$ and 1 GeV, and a brane thickness of $L^{-1}=2$ TeV, the latter value chosen to avoid LHC constraints. Taking these radii, we have set  conservative lower or upper bounds on the bulk Dirac and Majorana  masses, the BLKT parameter ($\delta_A\alpha$) and the 4-D Yukawa couplings $\lambda_{4, 1(2)}$. These examples are representative for other plausible values of parameters that would give similar results. The most favorable compactification radius is $R^{-1}=1$ GeV because it allows 4-D Yukawa couplings of order one for a mass of the lightest sterile neutrino of order $30$~TeV, while for the other radii the couplings must be smaller.  Masses of the lightest sterile neutrinos  of order $\mathcal{O}(1-10)$ \text{TeV} easily satisfy the required SM neutrino mass if the 4-D Yukawa couplings are some orders of magnitude smaller, such as $10^{-2}-10^{-1}$.

In addition,  neutrino oscillation experiments do not impose challenges to the model, because the SM neutrino practically does not oscillate into any other sterile neutrino KK state. The present setup can also explain the observed baryon asymmetry of the Universe through leptogenesis, but that prediction is not generic where smaller and larger values by order of magnitude can result from the plausible range of parameters. Additional fermions in the bulk, with different flavors, would give similar results and, although it is beyond the scope of the present work, atmospheric and solar neutrino mass splitting can easily be accommodated in this model and will be investigated in the future.  

Finally, potential signatures for this model include searches for UED particles, where the cascade decay of SM KK particles constrains the UED compactification radius $L$. Missing energy from additional KK states (from sterile neutrinos) may be expected to be seen along with UED KK particles, if their masses are low enough. In this case, it would be possible to infer the necessary 4-D Yukawa couplings to produce the seesaw mechanism.
    
\acknowledgments
R.G.L. acknowledges CAPES (process 88881.162206/2017-01) and Alexander von Humboldt Foundation for the financial support.

\bibliography{references}

\end{document}